\documentclass[a4paper,12pt]{article}

\usepackage[latin1]{inputenc}
\usepackage{fontenc}
\usepackage{graphicx}

\usepackage[dvips]{hyperref}

\begin{document}

\title{\bf Modelling survival and allele complementation in the evolution of genomes with polymorphic loci}
\maketitle

\bigskip
\noindent
S. Cebrat$^{1}$, D. Stauffer$^{2,3}$, J.S. S\'a Martins$^{1,2,4}$, S. Moss de Oliveira$^{1,2,4}$, and P.M.C. de Oliveira$^{1,2,4}$

\bigskip
\noindent
$^1$ Department of Genomics, Wroc{\l}aw University, ul. Przybyszewskiego 63/77,
51-148 Wroc{\l}aw, Poland 

\medskip
\noindent
$^2$ Laboratoire PMMH, \'Ecole Sup\'erieure de Physique et de Chimie
Industrielles, 10 rue Vauquelin, F-75231 Paris, France

\medskip
\noindent
$^3$ Visiting from Institute for Theoretical Physics, Cologne University,
D-50923 K\"oln, Euroland

\medskip                
\noindent
$^4$ Visiting from Instituto de F\'{\i}sica, Universidade Federal Fluminense, and National Institute of Science and Technology for Complex Systems; Av. Litor\^{a}nea s/n, Boa Viagem,
Niter\'{o}i 24210-340, RJ, Brazil

\bigskip

Abstract

{\small 
We have simulated the evolution of sexually reproducing populations composed of individuals represented by diploid genomes. A series of eight bits formed an allele occupying one of 128 loci of one haploid genome (chromosome). The environment required a specific activity of each locus, this being the sum of the activities of both alleles located at the corresponding loci on two chromosomes. This activity is represented by the number of bits set to zero. In a constant environment the best fitted individuals were homozygous with alleles' activities corresponding to half of the environment requirement for a locus (in diploid genome two alleles at corresponding loci produced a proper activity). Changing the  environment under a relatively low recombination rate promotes generation of more polymorphic alleles. In the heterozygous loci, alleles of different activities complement each other fulfilling the environment requirements. Nevertheless, the genetic pool of populations evolves in the direction of a very restricted number of complementing haplotypes and a fast changing environment kills the population. If simulations start with all loci heterozygous, they stay heterozygous for a long time. 
}

\bigskip

\section{Introduction}

A Mendelian population is, by definition, a group of interbreeding diploid individuals sharing the same genetic pool. Such population should be panmictic, which means that each individual can randomly choose a mating partner from the whole population. If the population is large, one can expect random and independent assortment of alleles to the gametes. In fact, in Nature, populations of one species usually do not fulfil the parameters of the Mendelian population even if they occupy the same and uniform environment. First of all, inbreeding populations could be much smaller than the whole considered populations, not necessarily because of real physical or biological barriers, but simply because of physical distances between individuals. Wright has introduced the definition of effective population size \cite{Wright}.  According to this definition the effective population is ``the number of breeding individuals in an idealized population that would show the same amount of dispersion of allele frequencies under random genetic drift or the same amount of inbreeding as the population under consideration''.

Introducing the finite effective size of population, one has to consider much deeper changes in the population evolution. The most important is that in such populations some genes cannot be inherited independently of each other. Genes in the genomes are arranged linearly in the large sequences called chromosomes. A diploid genome is composed of several pairs of homologous chromosomes (i.e. the human genome is composed of 23 pairs of chromosomes). Even if we assume that chromosomes are inherited independently, the genes located on single chromosomes are genetically linked. Two homologous chromosomes (one inherited from the mother, the second from the father) can recombine during gamete formation exchanging corresponding parts, in a process called crossover, but the frequency with which this process occurs is restricted and relatively low. For example, the human chromosome 19 contains almost 1500 genes while the crossover happens usually only once during gamete production \cite{Jensen}, \cite{Kong}. As a result, two neighbouring genes can be inherited together as a linked unit for hundreds of generations. Thus, single genes are not inherited independently, they tend rather to form clusters of genes.

Deleterious genes can be even lethal, they can kill their carriers. Nevertheless, they can be compensated by a functional copy of the other allele located in the corresponding locus of the homologous chromosome since they are usually recessive. In situations of large Mendelian populations, deleterious genes are eliminated by purifying selection: If a given locus in a genome is occupied by two lethal alleles, the individual is eliminated by selection, thus decreasing the number of deleterious genes in the genetic pool. In cases where we should consider the inheritance of large clusters of genes, the strategy of population evolution could be different. Not only single deleterious genes can be complemented by their functional alleles, but whole clusters, with several deleterious genes, can be compensated by the corresponding fragment of the homologous chromosome. This strategy has been called the complementing strategy \cite{Zawierta}, as an alternative for the above purifying selection. In the complementing regions, the fraction of heterozygous loci (occupied by two different alleles) is much higher than in the regions under purifying selection and can reach up to $100\%$ of the positions \cite{Zawierta2}, \cite{Waga}, \cite{Waga2}. The transition between the two alternative strategies may have the character of a phase transition and it depends on the frequency of crossover between homologous chromosomes, effective population size, and the number of genes in the crossing chromosomes: Lower recombination rate, smaller effective population size, longer chromosomes and changing environment favour the complementing strategy \cite{Kowalski}, \cite{Zawierta2}. Complementing strategy allows sympatric speciation - emerging new species inside the larger ones without any barriers \cite{Waga}, \cite{Waga2}. This is possible because the complementing clusters in the modelled populations have a unique sequence of defective and wild alleles. Surviving offspring can be produced mainly by individuals sharing the same sequence of defective genes in the complementing clusters. On one hand, this strategy explains also the results showing that fecundity in the human population decreases with large genetic distance between spouses \cite{Helgason}. According to the predictions of models of population evolution under purifying selection, on the other hand, fecundity should increase with the genetic distance between spouses, while models with  gene cluster complementation predict both, inbreeding and outbreeding depression \cite{Stauffer}, \cite{Stauffer2}. The last predictions of the models are in agreement with other findings of Helgason et al that evolutionary success measured in the number of grandchildren has a maximum for spouses which are the third level of cousins.

In the previous papers dealing with complementing strategy only lethal characters of defective alleles have been considered. Thus, each gene could be represented only by two different alleles: the wild one, which is functional, and the defective one, lethal in the homozygous state and neutral in the heterozygous state. Many agent-based models of population evolution share this feature \cite{newbook}. In this paper we are introducing more polymorphic alleles with hundreds of possibilities (morphs). The environment requires specific values of loci activities and these requirements may change during the evolution of the population \cite{anais}. The required activity can be realised by several combinations of different alleles of a specific gene. Since a single allele is now composed of eight bits, the number $256$ of different alleles which can occupy a given locus is high and the population can be highly polymorphic.

\section{The model}

A diploid individual genome is represented by a pair of parallel bit-strings, each one with 1024 bits. Each string is divided into $L=128$ equal pieces or loci of 8 bits each. Thus, each genome consists of a pair of homologous chromosomes with 128 pairs of alleles located at corresponding loci. Each string corresponds to one chromosome and each locus of 8 bits is occupied by one allele. The two alleles at the corresponding loci on two chromosomes are denoted $A$ and $a$, but this terminology does not indicate a dominant or recessive character of these alleles. We denote as activities $A_l$ and $a_l$, with $l = 1,2 \dots L$, the number of bits set to zero in the corresponding alleles. The activity of the $l$-th locus is given by the sum $A_l + a_l$, which is a number between 0 and 16.  

The environment is also represented by a (single) string of $L=128$ loci, where the value $E_l$ 
of each locus represents the ideal activity of the corresponding $l$-th locus of each genome. The total deviation $D=\sum_{l=1}^L|A_l+a_l-E_l|$ determines the individual survival probability $x^{1+D}$ during one iteration, with a free parameter $x$ slightly below 1. With this selection strategy even perfectly fitted individuals have a probability $1 - x$ to die at each iteration.

For each adult who dies, one new baby is born in the same iteration, from two randomly selected different partners; no distinction between males and females is made. For each of the two partners, the genome strings are crossed over with an intergenic recombination probability $C \le 1$; thus the 8 bits of each locus are always kept together. The baby gets with probability $m$ one new mutation in each of its two genome strings, modelled by inverting the value of one randomly chosen bit. If pre-natal selection is used (in section 4 but not in section 3), the baby has to pass the above selection mechanism, depending on its deviation $D$, in order to be born; if it is not born, a new pair of parents is randomly selected and a new trial is made, up to some maximum number of attempts. The population size $N$ is fixed if pre-natal selection is not used, but we regard the population as extinct and stop that simulation if the number of surviving adults, without the newborn babies, approaches zero. With pre-natal selection, the population size may decrease if the maximum number of trials for a newborn is reached without any success in passing the selection mechanism. The environment string may change at each iteration with a low probability $e$ by changing one $E_l$ by $\pm 1$ at a randomly selected locus $l$.

As mentioned in the introduction, our main purpose is to check if this more realistic model also presents the two different strategies of evolution, the Darwinian purifying selection and the complementarity of haplotypes, and how these patterns behave in a changing environment. For this purpose we first need surviving populations. It is important to notice that in this model complementarity of haplotypes at a given locus means that the activities of the two corresponding alleles are different (heterozygous locus), but their sum is close enough to the environmental requirement for that locus for the individual to stay alive. On the other hand, purifying selection favours the homozygous state of loci, with the same activities of both alleles. Because we failed to reach full complementarity (for all loci) starting from  homozygous loci, we decided to first start from a fully complementary configuration and check its stability, as described in the next section. Results for other initial configurations will be discussed in section 4.

\section {Starting from full complementarity}

Initially we set $E_l = 8$ for all loci, meaning that the ideal activity (sum of the activities of both alleles) of any locus equals 8, and set all alleles of one bitstring to have bits 00000000 while all alleles of the second bitstring have bits 11111111. That is, we start with ideal complementary activities and check their evolution. Our parameters are: population size $N = 5000$, number of genes $L=128$, selection strengths $x$ = 0.9 (strong selection) or $x = 0.995$ (weak selection), no pre-natal selection, probability of changing one locus of the environment string $e = 10^{-4}$ (slowly) or $e = 10^{-2}$ (rapidly changing), total number of time-steps between 0.1 million and 10 millions and mutation probability per string at birth $m=0.9$, all reversible ($0 \leftrightarrow 1$).

At each time step and for each adult individual the activities $A_l$ and $a_l$ are computed and their differences $\Delta = \sum_{l=1}^L|A_l-a_l|$ and deviations from environmental ideal $D=\sum_{l=1}^L|A_l+a_l-E_l|$ are calculated. 
In the following figures we show the evolution of the averages per locus of both quantities, $<\Delta>/L$ and $<D>/L$. Due to the initial condition, we always have $<\Delta>/L$ starting from eight and $<D>/L$ starting from zero.

Fig.1 shows for $x=0.9$ and a constant environment that for $C=1$ both averages rapidly approach the value 4, while for $C=0.512$, $<\Delta>/L$ decays from 8 to 5 and $<D>/L$ increases from zero to 3. 
Such results indicate that the initial heterozygosity is maintained, although with fluctuations which are stronger for larger $C$, since crossing mixes the genomes disturbing their initial perfect heterozygous configurations. (We will turn back to the value 4 mentioned above when discussing Fig.4.)

\begin{figure}[!hbt]
\begin{center}
\includegraphics[scale=0.5,angle=-90]{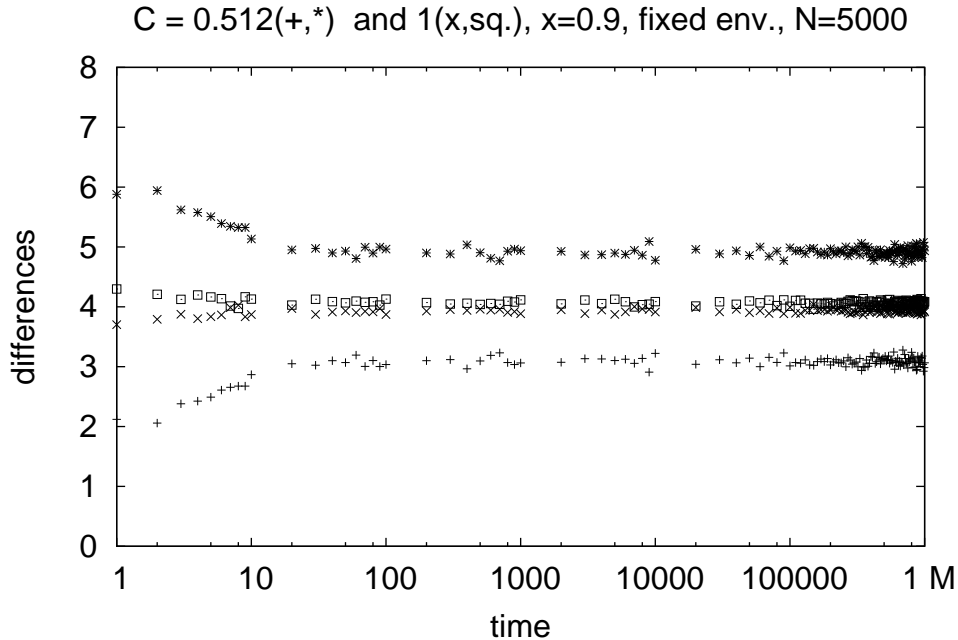}
\end{center}
\caption{Constant environment. Average difference (per locus) between the two genome strings (upper curves) and between genome and environmental requirements (lower curves). $C$ decreases from inside out.}
\end{figure}

Fig.2 shows the results for a slowly ($e=10^{-4}$, part a) and a more rapidly ($e=10^{-2}$, part b) changing environment, also including the crossing probability $C=0.256$. Now both averages approach the value 4, independently of the $C$ value. Notice the changed time scale from part (a) to part (b) due to early extinction: for $x=0.9$ populations do not survive in a drastically changed environment. 
Note that for $e=0.01$ we have one unit, $E_l$, of the environment string mutated at every 100 steps. 
After 10,000 steps on average 100 units were mutated, meaning that most of the 128 ideal gene activities were changed. If selection pressure is high, there is no time for the population to adapt to these changes. For $e=0.0001$ the population died out only after almost 1 million time steps.

\begin{figure}[!hbt]
\begin{center}
\includegraphics[scale=0.45,angle=-90]{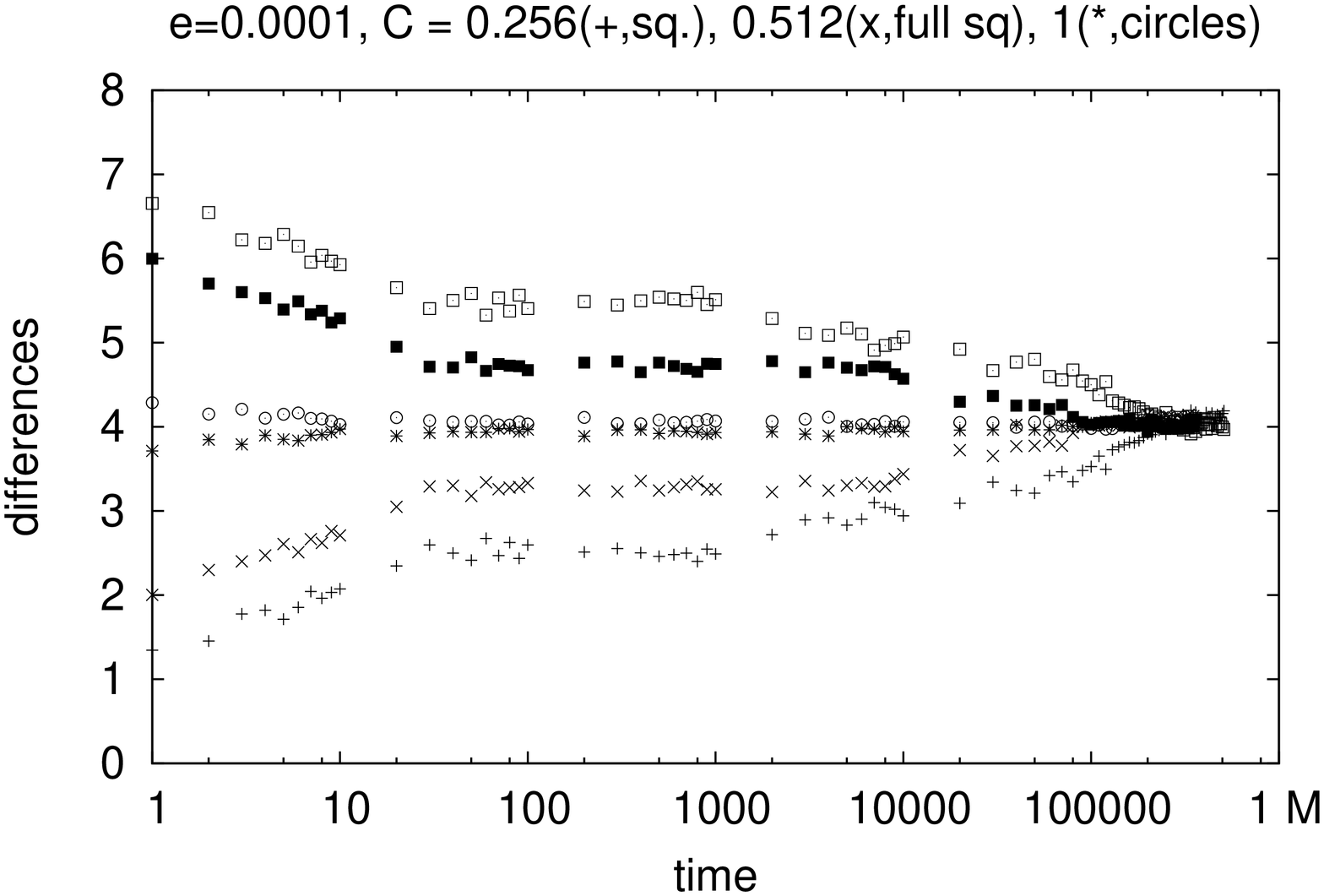}
\includegraphics[scale=0.45,angle=-90]{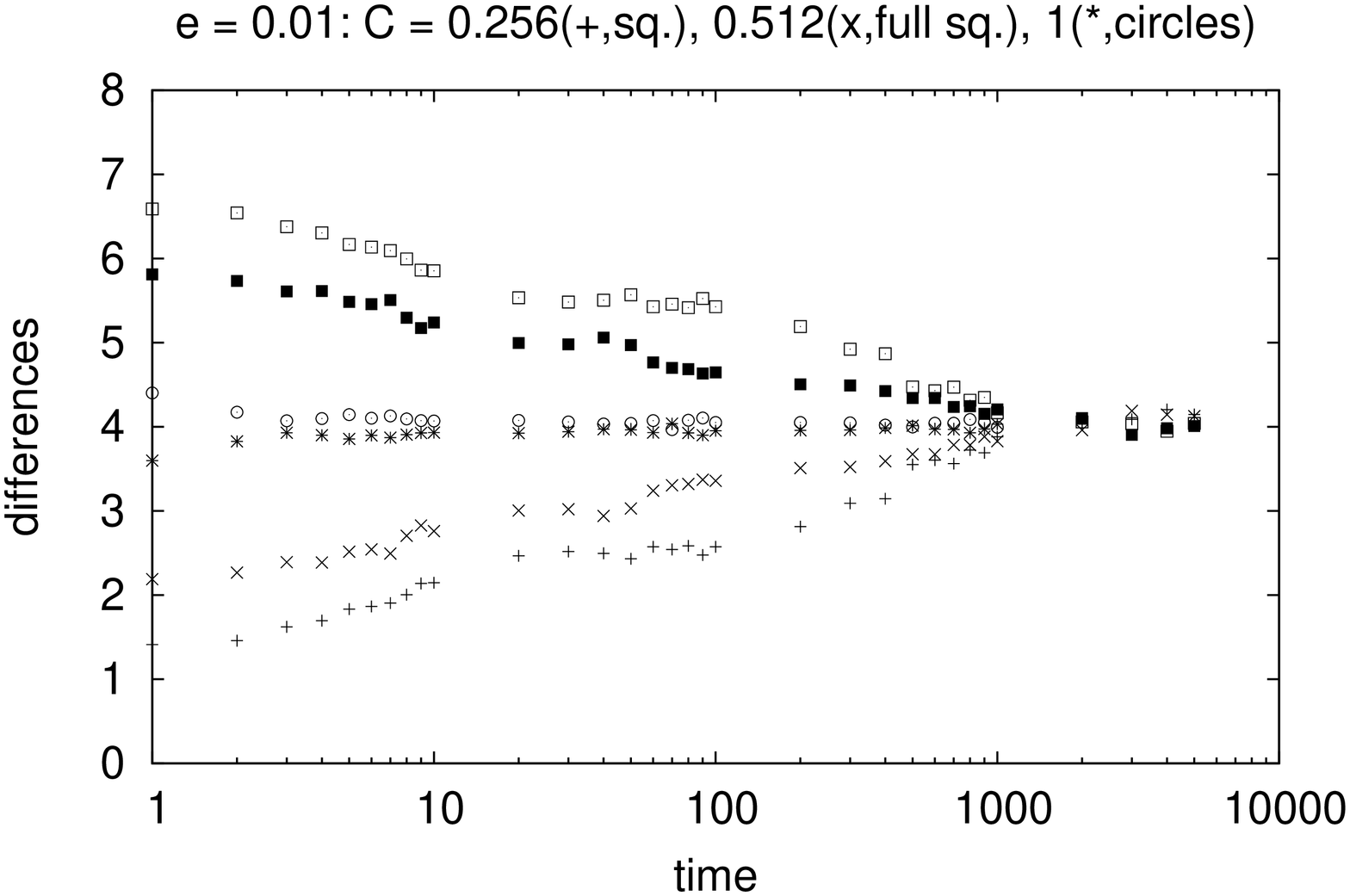}
\end{center}
\caption{As Fig.1, but for changing environment. Part (a): $e=0.0001$; Part (b): $e=0.01$. 
Now $C = 0.256$ (outermost curves) included, besides $C=0.512$ and 1. In part (b) population 
dies out much earlier, also if $N$ is increased from 5,000 to 50,000, 500,000 and 5 million.}
\end{figure}

Fig.3 shows that if the selection pressure is reduced, $x=0.995$, the populations are stable for times above one million time-steps even for a rapidly changing environment (part a) and above 10 million time-steps (part b) for a slowly varying one, even for $C=1$. 

\begin{figure}[!hbt]
\begin{center}
\includegraphics[scale=0.45,angle=-90]{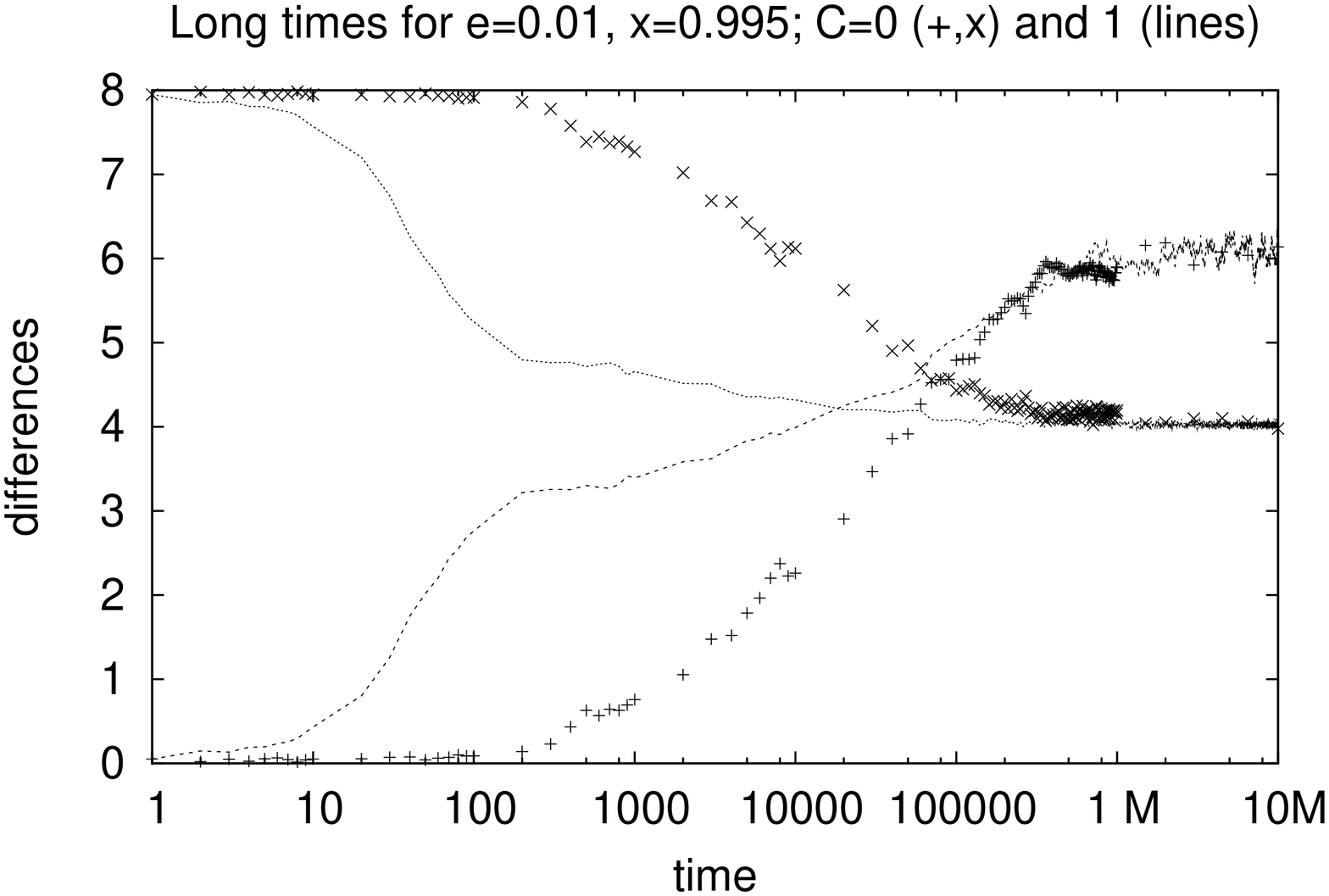}
\includegraphics[scale=0.45,angle=-90]{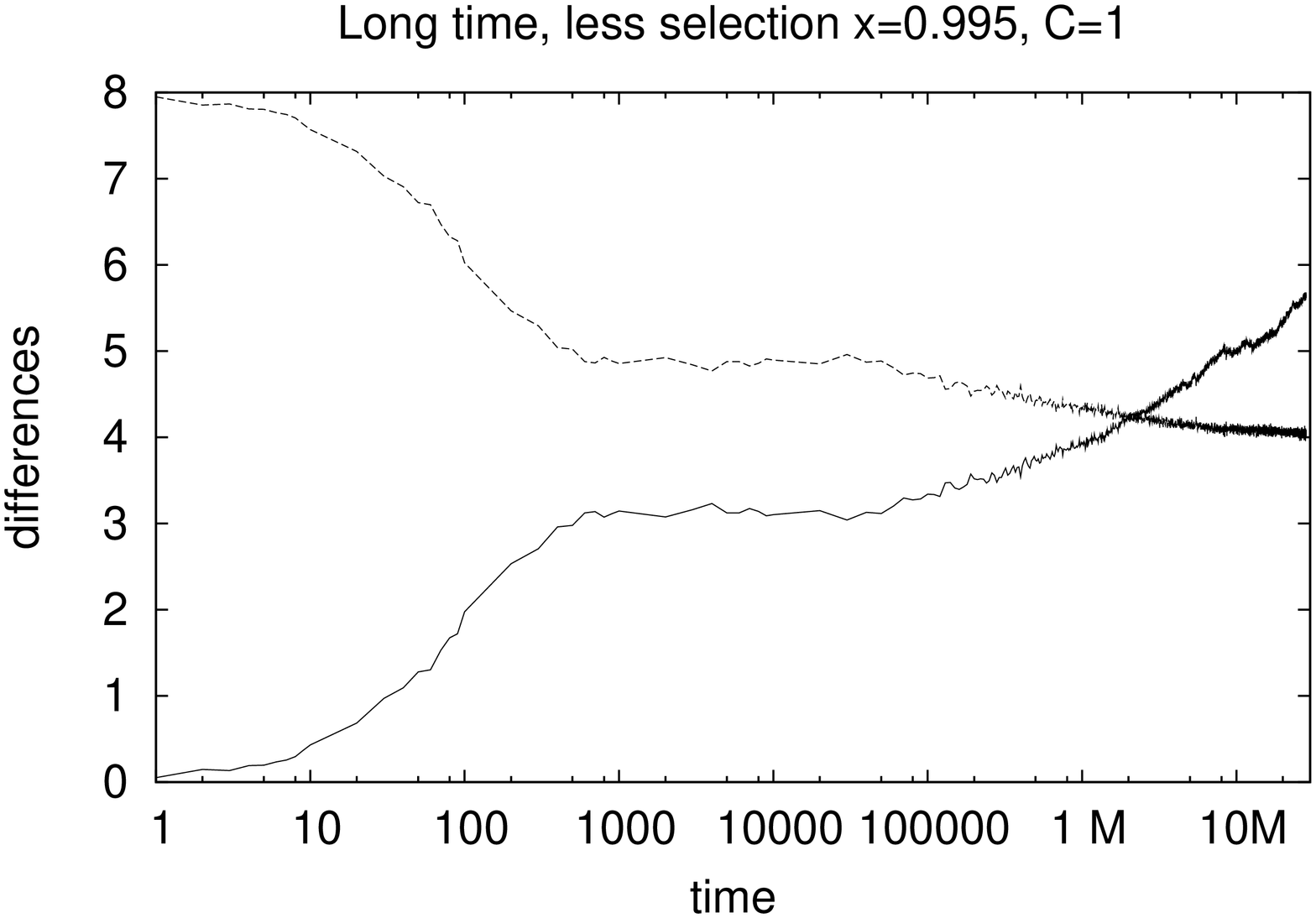}
\end{center}
\caption{Weaker selection, $x=0.995$. In this case populations may survive much longer. 
Part (a): $e=0.01$; Part (b): $e=0.0001$.}
\end{figure}

Now we go back to $x=0.9$ in constant environment and present in Fig.4a the histograms for $\mid A_l-a_l \mid$ and for the bit-by-bit Hamming distance between the corresponding 8-bit genes of the two genome strings. For $C=1$ about half of these distances are 0 and half are maximal (8), which explains why the averages $<\Delta>/L$ presented in Fig.1 go to 4. Fig.4b shows the variation with the recombination rate $C$. (For these histograms all time steps between 90,000 and 100,000 and all adults were summed over.)  

\begin{figure}[!hbt]
\begin{center}
\includegraphics[scale=0.45,angle=-90]{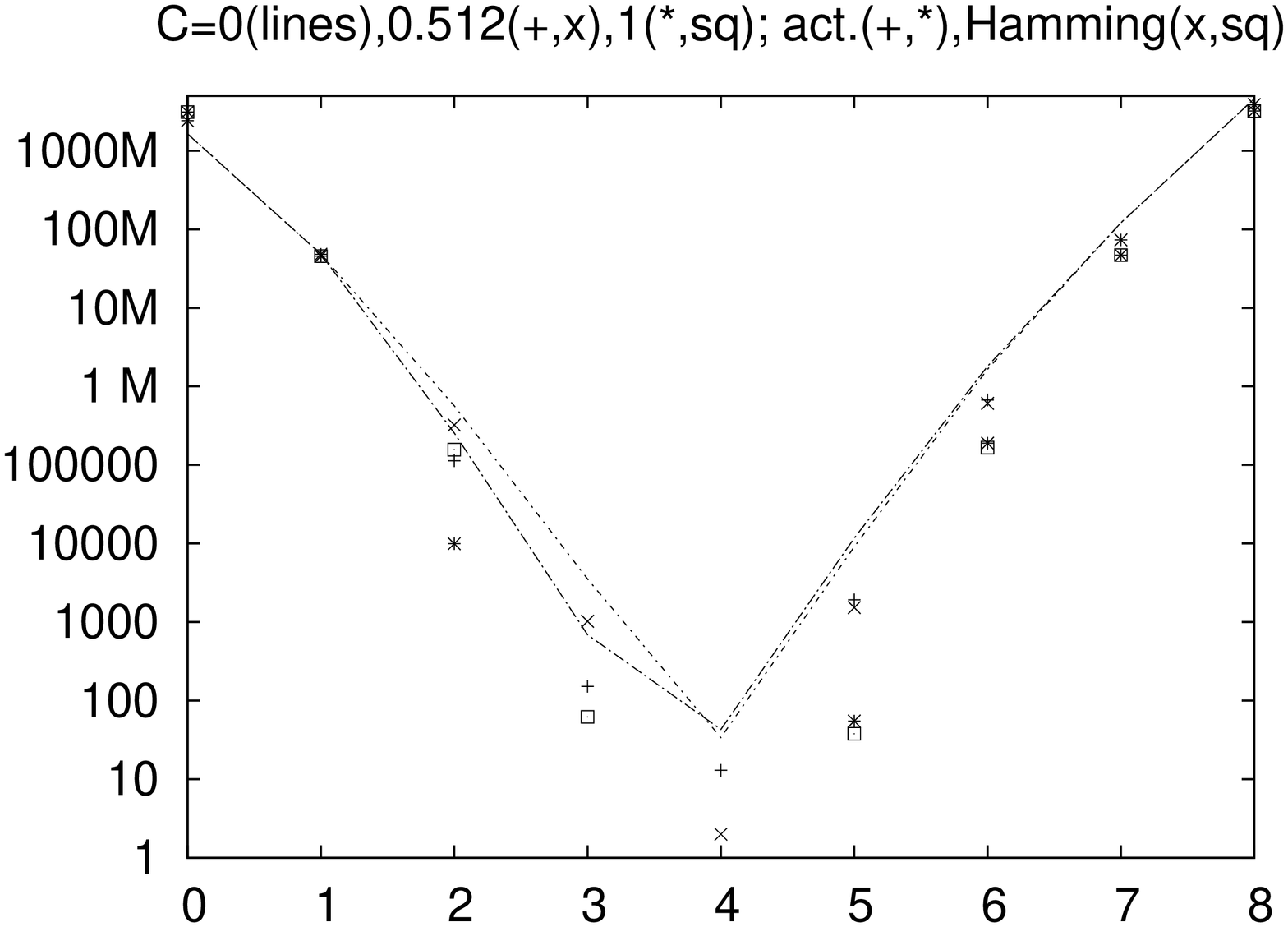}
\includegraphics[scale=0.45,angle=-90]{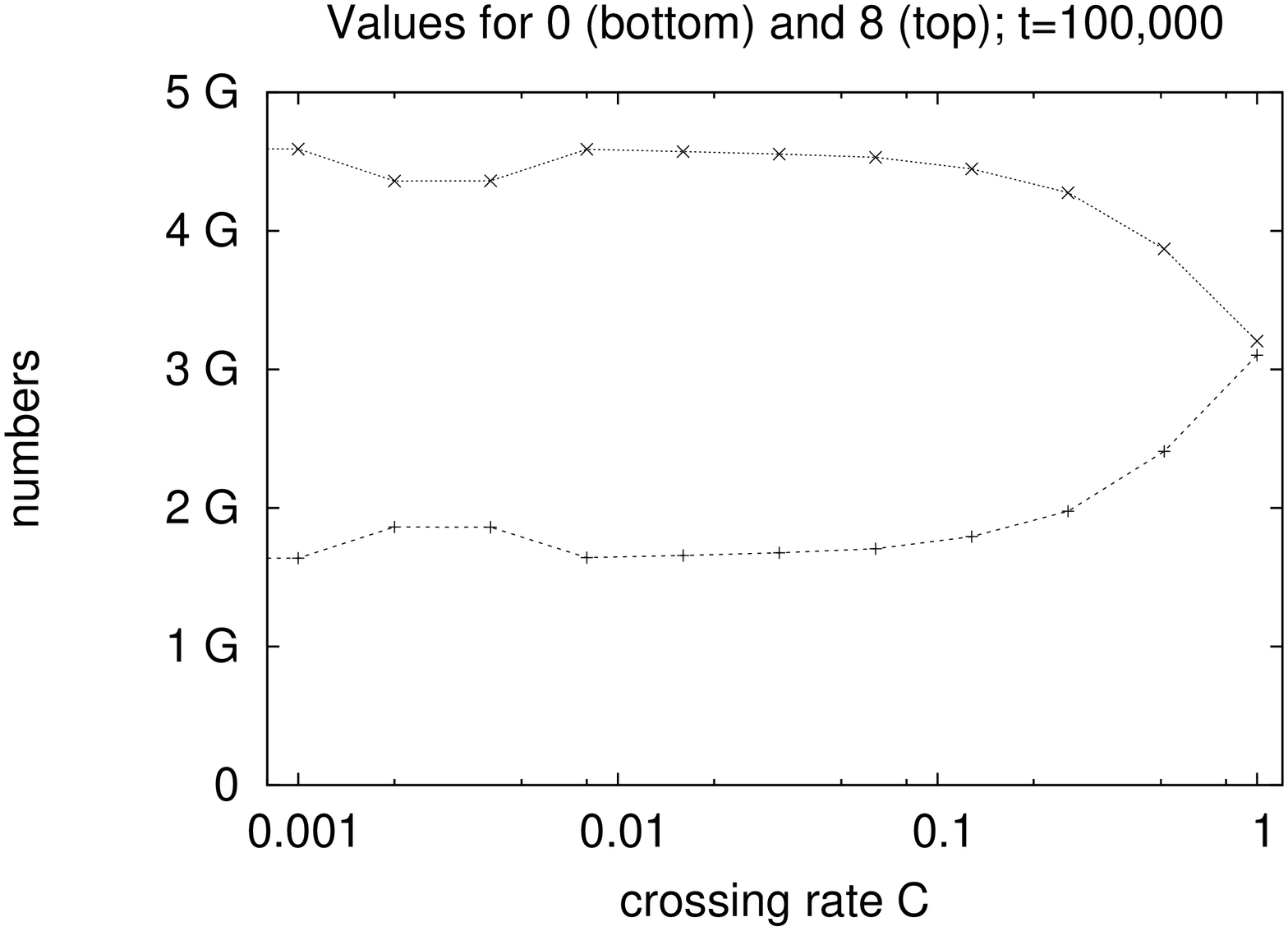}
\end{center}
\caption{Constant environment, $x=0.9$. Part a: Histograms for the activity difference (+,*) and the bit-wise Hamming distance (x,squares); for $C=0$ (lines) the results are barely distinguishable. Part b: Histogram values at 0 and 8 versus $C$; differences (symbols) and Hamming distances (lines) again barely differ. Parameters as in Fig.1; $e = 0$.}
\end{figure}

In all results presented so far averages were performed at each time-step before applying the environment-dependent selection mechanism. If they are performed after selection, the populations are much smaller (since only adults contribute to them; not shown), and the averages stay much closer to the initial full complementarity, as shown in Fig.5.

\begin{figure}[!hbt]
\begin{center}
\includegraphics[scale=0.45,angle=-90]{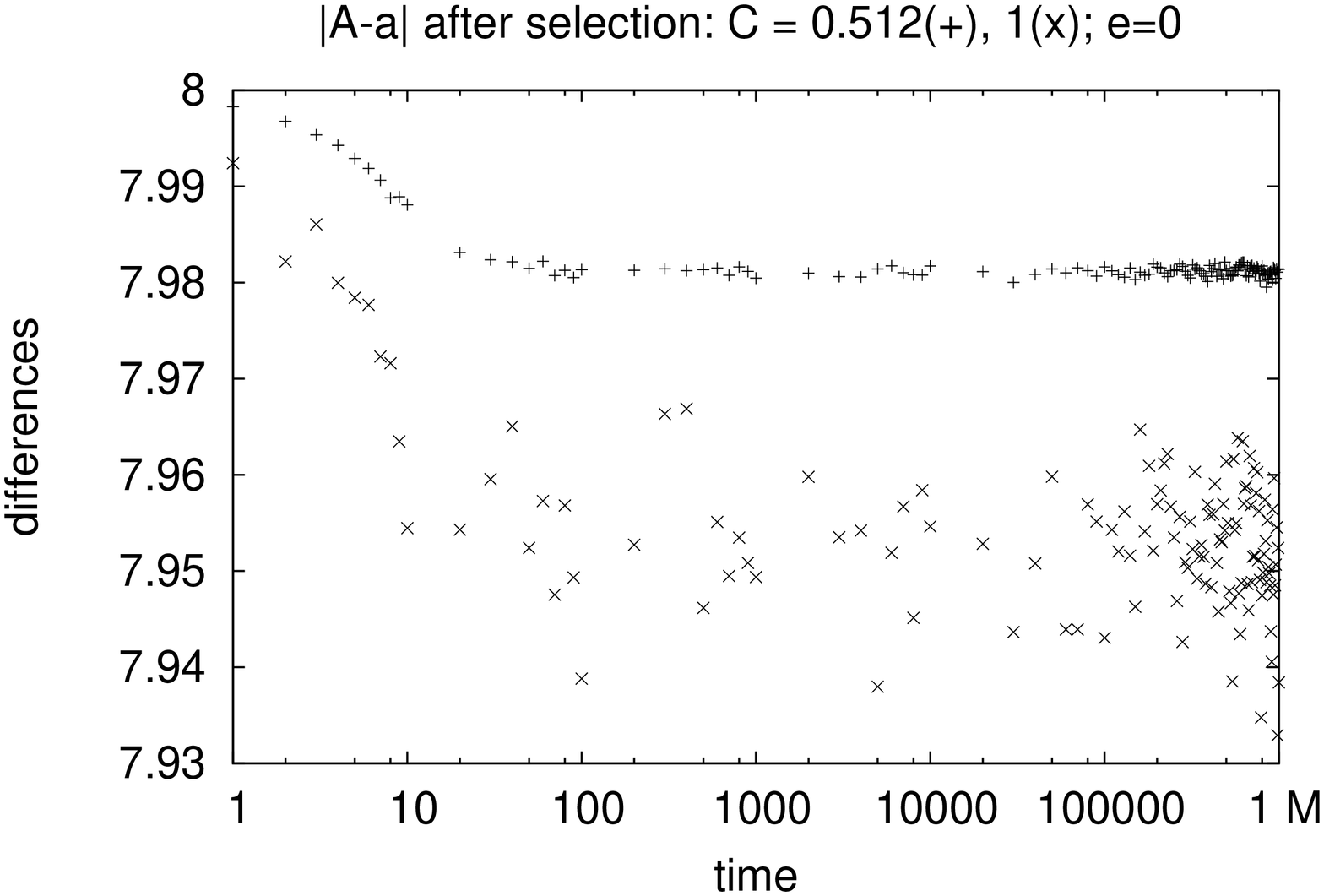}
\includegraphics[scale=0.45,angle=-90]{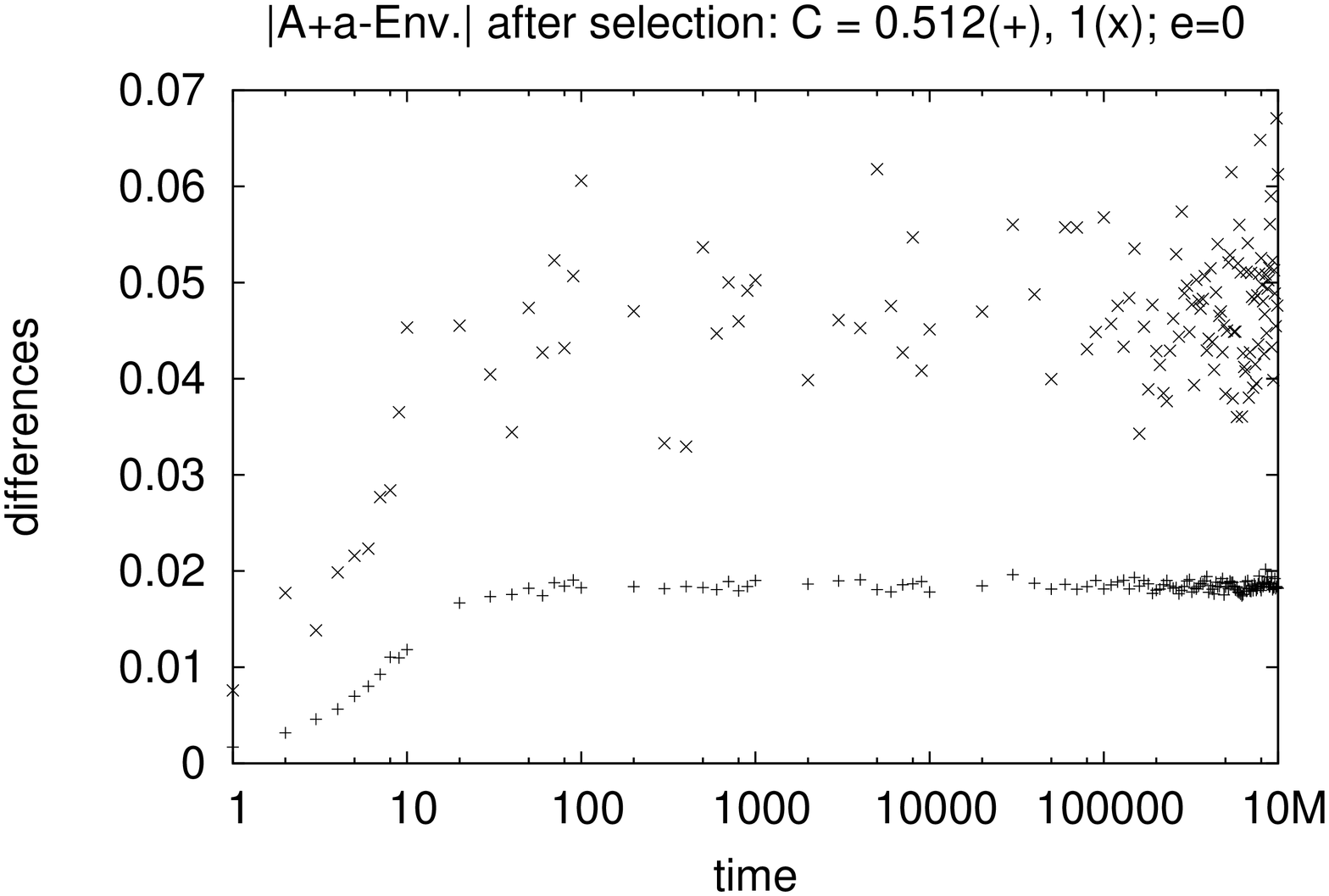}
\end{center}
\caption{Constant environment, averages performed after selection, $x=0.9$ for $C=0.512$ and $C=1$. Part a: activity differences; Part b: distance to environment.}
\end{figure}

Results don't change much if mutations are made irreversible instead of reversible, babies are born from adults only, the genome length $L$ is shortened from 128 to 32 (or lengthened to 512) or the mutation rate $m$ is lowered from 0.9 to 0.1 and 0.01.

\section{Other initial configurations}

When we started our simulations our intention was to check if we could reach full complementarity and not to depart from it. For this reason we established that the ideal number of bits set to $0$ at a locus would have to be even: an odd-valued requirement would favour heterozygosity at this locus, independent of the value of the crossing probability, since the number of bits set at any locus of a haplotype is an integer in the model.

A number of different initialisations were then tried, before we could find one that led the resulting genetic pool of the population to resemble the expected complementarity. Those attempts were:

1. The environment requires the same activity $E_l=8$ for each locus and all individuals have all loci in the homozygous state - at each locus, each haplotype has 4 bits, chosen randomly, set;

2. The environment requires the same activity $E_l=8$ for each locus and half of the individuals are in the homozygous state, as above, and the other half in the extreme complementary state, as in the previous section.

3. Initially the environment requires no bits to be set at all loci and all individuals are ideal, i.e. have no bits set. Then the environment changes constantly with some probability $e$ per step. 

The results of strategy 2 quickly decay into those for strategy 1, no matter what value of $C$ is used. This means that the complementary state is metastable and that the homozygous state is much more robust. In fact, when strategy 1 is used the population never leaves the homozygous state, for all values of $C$. These results may be understood by the following reasoning: in the perfectly heterozygous state, if the effect of new mutations is omitted, only 50\% of the tentative newborn will have complementary genomes while the other 50\% have genomes composed of two identical haplotypes, which are lethal. In the case of a full homozygous state, all newborns are surviving.

Strategy 3 has as a drawback the fact that if a mutation of the environment pattern adds $\pm 1$ to the required number of bits set at some locus the end result may be an odd number, and this 
does not allow that particular locus to be homozygous for a well fit individual. If we add or 
subtract two, then the load carried by each environment mutation is so strong that it may lead the 
population to extinction. There are, nevertheless, unexplored regions of parameter space that might 
lead to interesting developments using this strategy or some variant thereof.

The procedure that gave heterozygosity starts with the environment requiring no bits to be set at all loci and all individuals are ideal, i.e. have no bits set. The population is allowed to evolve for a number of time steps proportional to the initial population and then starts an equilibration procedure: at each Monte Carlo step, with a $0.001$ probability, the environment changes its requirement at some random locus, by increasing by one the number of set bits required. This procedure continues until $16$ loci require each of the even values $0, 2, ..., 14$, which happens, for the parameters we used, shortly before 1 million time steps have elapsed. The environment loci requirements are no longer increased during this initialisation after having reached these values.

\begin{figure}[!hbt]
\begin{center}
 \includegraphics[angle=-90,scale=0.5]{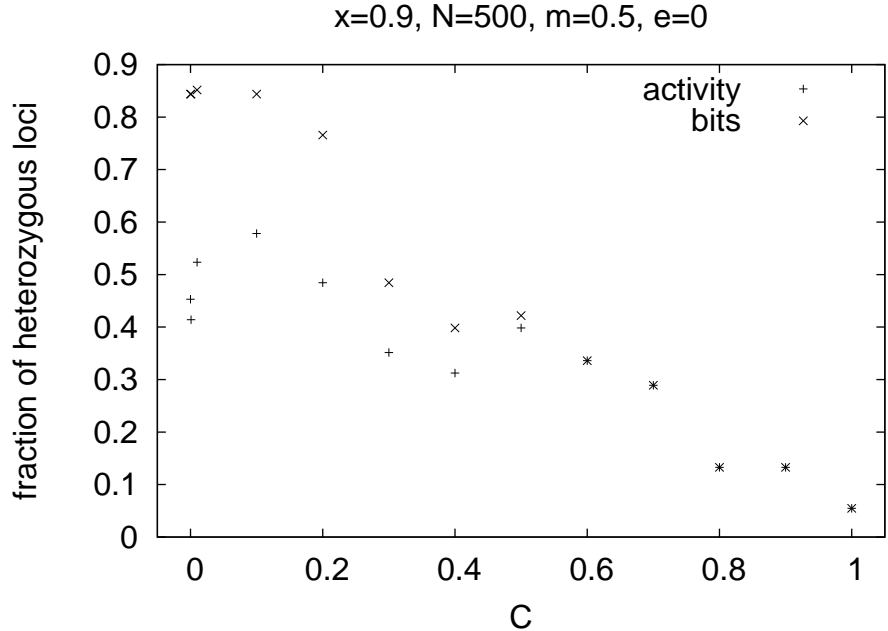}
 \newline
 \caption{Fraction of heterozygous loci as a function of $C$.\label{Heteloci}}
\end{center}
\end{figure}

We start presenting the most important results we got with this last version of the model, which also includes pre-natal selection, namely the partial answer to the original question about the establishment of a  purifying regime for large values of C and of a complementary one for small values of this parameter. Fig. \ref{Heteloci} shows this feature for two different measures of heterozygosity, which is now understood to be the translation of complementarity for this model. One is activity-wise: a locus is homozygous if the activity of both its alleles are the same, it is heterozygous otherwise. The fraction of the loci that are heterozygous in this concept is plotted against the value of $C$ (+ symbols). The other measure is bit-wise: a locus is homozygous if the bit  configurations of both its alleles are precisely the same, and are heterozygous otherwise. The fraction of loci that are heterozygous in this concept is shown as a function of $C$ ($\times$  symbols). The end result of both measures is similar: this fraction is higher for small than for large $C$. Nevertheless, the heterozygosity measured by bit-wise Hamming distance is higher, which means that polymorphic alleles of the same activity for the same locus were generated several times independently if the recombination rate was low. Under higher recombination rates, the fraction of heterozygous loci measured by both Hamming distances, activity-wise and bit-wise, are the same, suggesting other mechanisms of generating the polymorphism. The results shown were produced by one single run for each value of $C$, as in the previous section. The other parameters are shown in the figure title.

\begin{figure}[!hbt]
 \centering
 \includegraphics[angle=-90,scale=0.5]{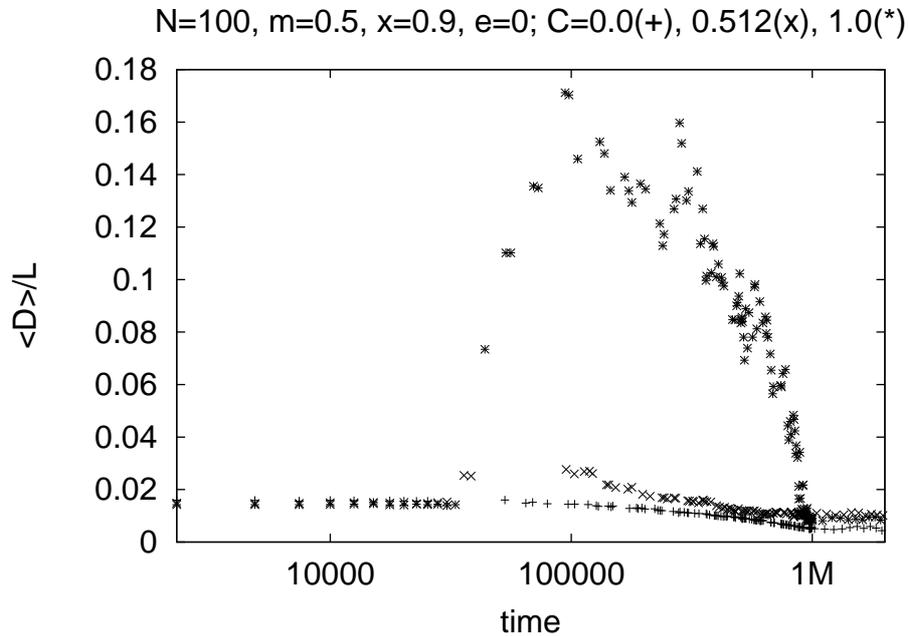}
 \newline
 \caption{Time evolution of the average deviation from the environment ideal (per locus).\label{fig:DxT}}
\end{figure}

Now come some time evolution plots. The first, Fig. \ref{fig:DxT}, shows how the average deviation $D$ from the environment ideal per locus evolves. The equilibration step of this version starts at step $64,000$ for the parameters used in this plot. Up to that point, the average deviation per locus, which started with a value of zero, grows thanks to the mutations to a stable value below $0.02$ for all values of $C$. For $C=0$ (no recombination) it stays small and even decreases during the equilibration process to half of its original value. For $C=0.512$ it increases during equilibration but not much, and by the end of this step it is back to the value it had before equilibration. For $C=1$ the behaviour is completely different: as equilibration starts, the deviation first increases up to nine times its original value, and can lead smaller populations to extinction. At the end of this step, the deviation decreases back to its starting value and at the end, when the population is brought down to the value of $N$, lies in between the values for the two other values of $C$. This difference between $C=0.512$ and $C=1$ agrees with Figs. 1, 2, 4b.

\begin{figure}[!hbt]
 \centering
 \includegraphics[angle=-90,scale=0.5]{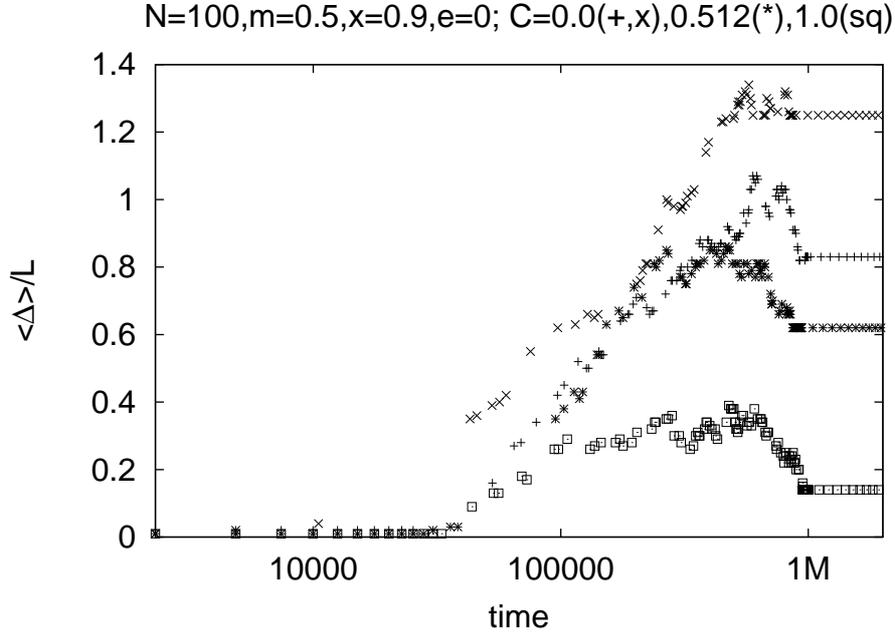}
 \newline
 \caption{Time evolution of allele-activity difference (per locus). Initial population is $2,000$ for all simulations except for $C=0.0$ ($\times$ symbol) where it is $300$.}
 \label{fig:AxT}
\end{figure}

Fig. \ref{fig:AxT} shows how the allele activity difference per locus $<\Delta> / L$ evolves with time. Before the start of the equilibration process it has a small residual value, reflecting the initialisation of the population and of the environment ideal (all zeroed for all loci). During the equilibration process it first rises for all values of $C$, more steeply for smaller values then for larger ones, but it decreases afterwards. At the end of this phase of the simulation it stabilises, albeit at values that depend on $C$ monotonically: they are larger when $C$ is smaller. This result reflects again the answer given by simulations of this model to the original question. For $C=1$ the population is close to being completely homozygous, which leads the difference between allele activities to a small value, while for $C=0$ heterozygosity emerges and the value gets close to unity. As in Figs. 1-7, homozygosity increases with increasing recombination.

The importance of inbreeding for this effect is shown by the data from a simulation where the population size is initially $300$, whereas this value was $2,000$ for all the others. Now, not only the size of this difference is bigger but also the fraction of loci where the allele difference is non-zero increases (not shown). This effect is even stronger for a population of $20,000$ and agrees with earlier simulations [4-7,10,11], but contradicts our results in section 3 (figure caption 2b) where $N$ had little influence.

\begin{figure}[!hbt]
\begin{center}
 \includegraphics[angle=-90, scale=0.5]{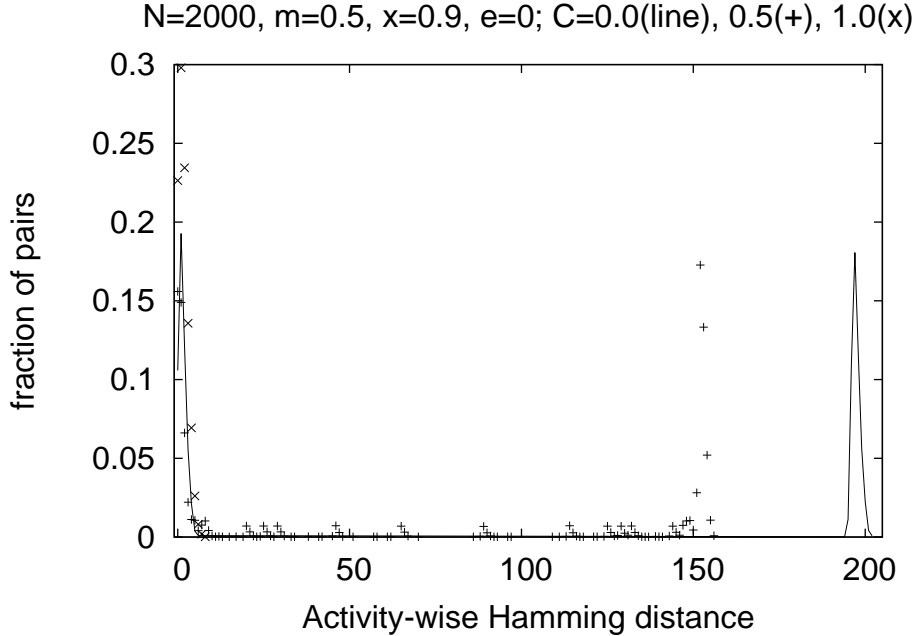}
\caption{Distribution of activity-wise Hamming distance for the near-complementarity state.
\label{ham}}
\end{center}
\end{figure}

As a final statement about the issue of purifying selection versus the strategy of complementarity, or homo- versus heterozygosity, we present our Fig. \ref{ham}. It shows the distribution of the activity-wise Hamming distance, defined as follows: for each individual of the population, each of its haplotypes is compared with each of the two haplotypes of all the other individuals, and the summed activity difference between them computed and normalised by the total number of such pairs. When $C=1$ this distribution has a single peak close to $0$. In fact, in this limit most loci are homozygous and the individuals are all very similar to each other. The resulting distances are all very small, and it does not matter which haplotype of one individual is compared to haplotypes of other individuals since they are essentially equal. This situation changes as $C$ decreases, and a double-peaked distribution develops, analogous to Fig. 4a. For $C=0$ it is clear that two patterns of haplotypes have been fixed in the population. Among each pattern the distances are very small, but the distance between two different patterns is substantial, characterising in a different way the establishment of a heterozygous regime in the population. For the strong selection that was used in the simulation, the average deviation from the ideal is essentially zero and the maximum value of this distance would then be $512$. The result of the simulation shows a peak close to $200$, at $0.4$ of the maximum.

\begin{figure}[!hbt]
 \centering
 \includegraphics[angle=-90,scale=0.5]{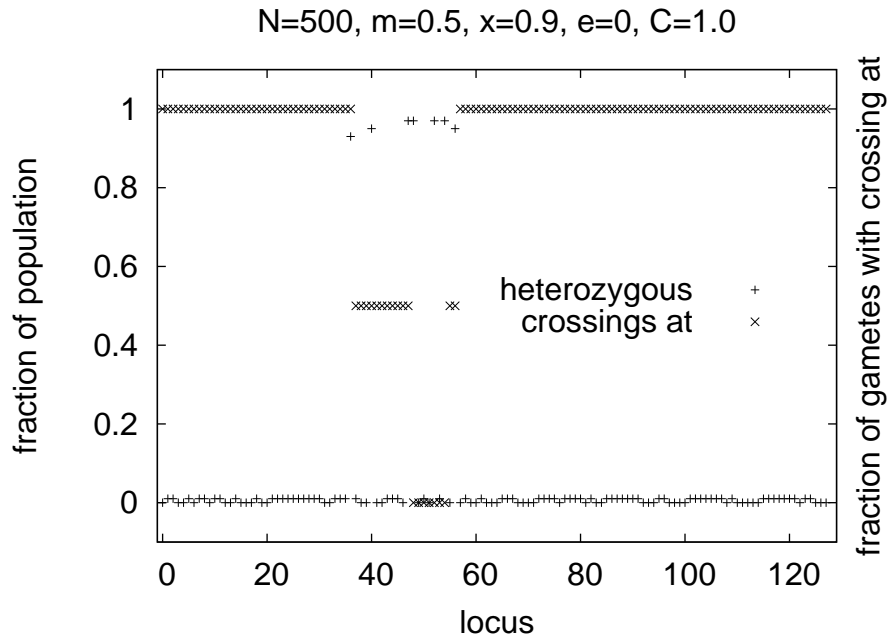}
 \newline
 \caption{Distribution of the crossing loci of the population and fraction of heterozygosity per locus.}
 \label{fig:Hetecros}
\end{figure}

This feature of high degree of homozygosity for $C=1$ has an interesting counterpart in the actual distribution of the crossing loci of the gametes of the population as shown in Fig.\ref{fig:Hetecros}. These loci are those points where the maternal and paternal bit-strings crossed over to produce those gametes which now form the investigated surviving adult. The few loci where full homozygosity did not set in are located close to the mid part of the chromosome - this is a general feature noticed in other models \cite{Waga,wroclaw} and it will be further discussed elsewhere. Because of this location, gametes generated by crossing near the middle of the chromosome will, with a very high probability, end up pairing into homozygosity for some of those loci and an activity far from the one required by the environment, leading the individual to fail pre-natal selection. The individuals that are alive come from crossings that avoid this region of the genetic material.

\begin{figure}[!hbt]
 \centering
 \includegraphics[angle=-90,scale=0.5]{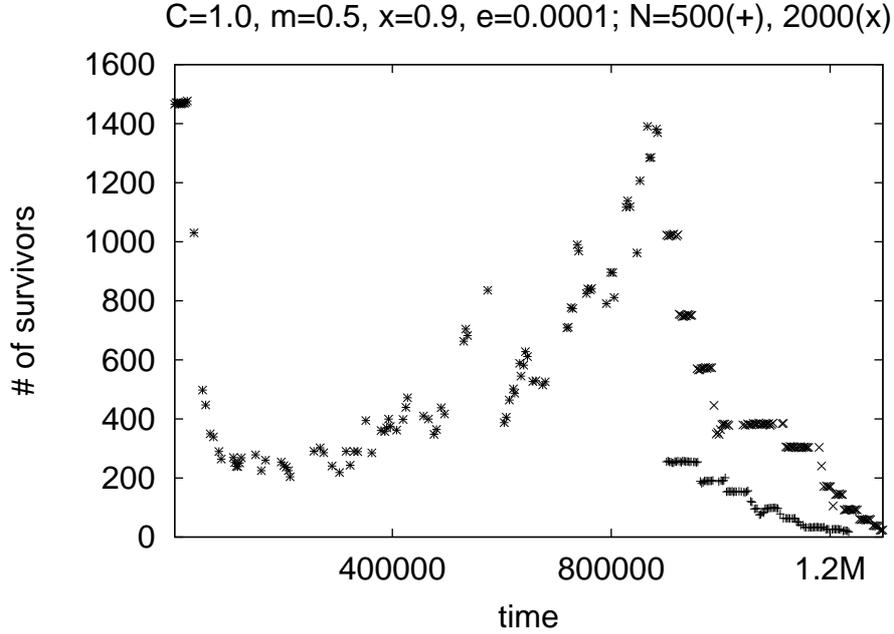}
 \newline
 \caption{Time evolution of the effective population size.}
 \label{fig:SxT}
\end{figure}

As a final remark, we show a time evolution plot for a simulation where the environment varies with some probability after the initial equilibration, Fig. \ref{fig:SxT}, while for Figs. 6-10 the environment was kept constant after its initial equilibration. It shows the effective population, defined as the difference between its actual size at the start of one Monte Carlo step and the number of individuals that die in that same step. For the case $C=1$ where we have maximum homozygosity in the population, this variability leads to a quick extinction. This feature has to be checked for other values of the parameters, in particular for a less selective environment.

\section{Discussion}

In this paper we have assumed that two alleles complement each other if they have different activity (heterozygous locus) and together fulfil the environment requirement. If both alleles have the same activity (homozygous locus), their relation is not considered as complementation even if they fulfil the environmental demands.

In a constant environment, the most robust populations have all loci homozygous. This is an obvious result because, independently of haplotypes' combination, the activities of loci corresponded exactly to the environment demand. Thus, ignoring new mutations, the fusion of gametes always resulted in forming a surviving newborn. If a specific locus can be occupied by alleles of different activities, only some of them combine in such a way that they fit exactly to the environment requirement. If we imagine such complementing alleles in a given locus, then to succeed in forming the fittest zygote, all other loci should also complement or should be homozygous. One should expect a tendency to form clusters of complementing alleles. This is more probable under low recombination rate and/or high inbreeding (smaller effective populations). In fact, we have observed this phenomenon (see Figs. 6 and 9) in our simulations. On the other hand, genes located at the ends of chromosomes, close to the telomeric regions, are more efficiently separated by recombination, while genes in the middle of chromosomes could be transferred into gametes as a linked unit more frequently. As a result, genes in the middle of chromosomes are more prone to form clusters of genes which are complementing other clusters. It is expected that in the regions of clusters, recombination is restricted. This is presented in Fig. 10,  even for relatively high recombination rate (C = 1). The same phenomena have been observed in models where genes were represented by single bits and existed in the genetic pool only into two states: wild-functional and defective-recessive lethal \cite{Zawierta}, \cite{Zawierta2}.

Parameters of the model presented in this paper seem to be too restrictive to allow the evolution of high polymorphism. The function of selection is too rigorous, killing even slightly unfit genomes. For the version of Section 4, the assumption that the activities of genes can be regulated in a broader range, like a survival probability exp$[-{\rm const}D^2]$ instead of $x^D$, should allow for the generation of a much higher polymorphism of the genetic pool, and it should be possible to find a configuration of parameters such that a larger set of haplotypes in the genetic pool could be more advantageous in a changing environment.


\begin{thebibliography}{88}
\bibitem{Wright} Wright, S., Evolution in Mendelian populations. Genetics {\bf16}, 97-159, 1931.

\bibitem{Jensen} Jensen-Seaman, M.I., Furey, T.S., Payseur, B.A., Lu, Y., Roskin, K.M., Chen, C.F., Thomas, M.A., Haussler, D., Jacob, H.J.  Comparative recombination rates in the rat, mouse, and human genomes. Genome Res. {\bf14}, 528-538, 2004.

\bibitem{Kong} Kong, A., Gudbjartsson, D.F., Sainz, J., Jonsdottir, G.M., Gudjonsson, S.A., Richardsson, B., Sigurdardottir, S., Barnard, J., Hallbeck, B., Masson, G., Shlien, A., P\'alsson, S.T., Frigge, M.L., Thorgeirsson, T.E., Gulcher, J.R., Stef\'ansson, K.  A high-resolution recombination map of the human genome. Nat. Genet. {\bf31}, 241-247, 2002.

\bibitem{Zawierta} Zawierta, M., Biecek, P., Waga, W., Cebrat, S. 2007. The role of intragenomic recombination rate in the evolution of population's genetic pool. Theory BioSci., {\bf125}, 123-132, 2007.

\bibitem{Zawierta2} Zawierta, M., Waga, W., Mackiewicz, D., Biecek, P., Cebrat, S., Phase Transition in Sexual Reproduction and Biological Evolution, Int. J. Mod. Physics C, {\bf19}, 917-926, 2008.

\bibitem{Waga} Waga, W., Mackiewicz, D., Zawierta, M., Cebrat, S., Sympatric speciation as intrinsic property of expanding populations, Theory in Biosciences, {\bf126}, 53-59, 2007.

\bibitem{Waga2} Waga, W., Zawierta, M., Kowalski, J., Cebrat, S.,  Darwinian purifying selection versus complementig strategy in Monte Carlo simulations From Genetics to Mathematics - Series on Advances in Mathematics for Applied Sciences, ed. J. Miekisz, M. Lachowicz, World Scientific. {\bf79}, 70-102, 2009.

\bibitem{Kowalski} Kowalski, J., Waga, W., Zawierta, M., Cebrat, S.,  Phase transition in the genome evolution favours non-random distribution of genes on chromosomes Int. J. Mod. Phys. C {\bf20}, 1299-1309, 2009.

\bibitem{Helgason} Helgason, A., P\'alsson, S., Guobjartsson, D.F., Kristj\'ansson, P., Stef\'ansson, K., An Association Between the Kinship and Fertility of Human Couples. Science {\bf319}, 813-816, 2008.

\bibitem{Stauffer} Bo\'nkowska, K., Kula, M., Cebrat, S., Stauffer, D.,  Inbreeding and outbreeding depressions in the Penna model as a result of crossover frequency. Int. J. Mod. Phys. C {\bf18}, 1329-1338, 2007.

\bibitem{Stauffer2} Cebrat, S., Stauffer D.,  Gamete recognition and complementary haplotypes in sexual Penna ageing model. International Journal of Modern Physics C {\bf19} 259-265, 2008.

\bibitem{newbook}
Stauffer, D., Moss de Oliveira, S., de Oliveira, P.M.C., S\'a Martins, J.S.,
{Biology, Sociology, Geology by Computational Physicists}. Amsterdam:
Elsevier 2006.

\bibitem{anais}
S\'a Martins, J.S., Stauffer, D., de Oliveira, P.M.C., Moss de Oliveira, S.,
Simulated self-organisation of death by inherited mutations. Anais Acad.
Bras. Cien., in press.

\bibitem{wroclaw}
D. Mackiewicz, M. Zawierta, W. Waga, S. Cebrat, Genome analyses and modelling the relationships between the coding density, recombination rate and chromosome length. preprint (2009).

\end{thebibliography}
\end{document}